\documentstyle[stwol]{article}

\def\Journal#1#2#3#4{{#1} {\bf #2}, #3 (#4)}

\def\NPB{{\em Nucl. Phys.} B}
\def\PLB{{\em Phys. Lett.}  B}

\def\be{\begin{equation}}
\def\ee{\end{equation}}
\def\bea{\begin{eqnarray}}
\def\eea{\end{eqnarray}}

\begin{document}

\title{STANDARD MODEL COUPLINGS WITH
{\bf\boldmath$\alpha^{-1}=137 \pm 9$}
FROM MULTIPLE POINT CRITICALITY AND 3 GENERATIONS OF GAUGE BOSONS;
NO GUT, NO SUSY}

\author{ D.L. BENNETT, H.B. NIELSEN }

\address{ The Niels Bohr Institute,
Copenhagen {\O}, Denmark}

\author{  C.D. FROGGATT }

\address{ Department of Physics and Astronomy,
 Glasgow University, Glasgow G12 8QQ, Scotland}


\twocolumn[\maketitle\abstracts{
We consider the application of the Multiple
Point Criticality Principle to the Standard Model
and its extension to the anti-grand gauge group
$SMG^3 \otimes U(1)_f$ at the Planck scale.
In particular we discuss the predictions for the fine
structure constants: $\alpha^{-1}(0) = 137 \pm 9$,
$\alpha^{-1}_{1}(M_Z) =99 \pm 5$,
$\alpha^{-1}_{2}(M_Z) =29 \pm 6$,
$\alpha^{-1}_{3}(M_Z) =12 \pm 6$.
We also present our Standard Model predictions for
the top quark and Higgs masses and a
$SMG^3 \otimes U(1)_f$ fit of the quark-lepton
masses and mixing angles. We characterise
$SMG^3 \otimes U(1)_f$ as the largest
possible extension of the Standard Model
gauge group satisfying some relatively simple
assumptions. The resolution of some other
fine-tuning problems ($\Lambda_{cosmological}$,
$\Theta_{QCD}$, $M_{Higgs}$) by the multiple point
requirement of degenerate vacua is briefly
discussed.}]

\section{Introduction}\label{sec:intro}

\vspace{-.2cm}

Using the multiple point criticality principle (MPCP)~\cite{db1,db2,db3}
together with the assumption
that the Standard Model Group (SMG) arises as the symmetry that survives the
Planck scale breakdown of a more fundamental gauge group
taken to be the 3-fold Cartesian product
of the usual SMG (denoted $SMG^3$),
we predict the Planck scale values of the Standard Model (SM)
gauge couplings with remarkable accuracy:
\begin{equation}
\begin{array}{lll}
   & \mbox{predicted} & \mbox{experimental} \\
\alpha^{-1}_3(M_Z) & 12\pm 6 & 9.25\pm 0.43 \\
\alpha^{-1}_2(M_Z) & 29\pm 6 & 30.10\pm 0.23 \\
\alpha^{-1}_1(M_Z) & 99\pm 5 & 98.70\pm 0.23 \\
\mbox{{\bf\boldmath $\alpha^{-1}$(0)}} & {\bf 137\pm 9} & {\bf 137.036...}\\
\end{array}
\end{equation}

The MPCP, which expresses the tendency for Nature to assume parameter values
corresponding to a maximally degenerate vacuum, can explain fine-tuned
parameters in general. However, the proposed underlying mechanism
implies
a (phenomenlogically tolerable~\cite{frogniel}) form of
non-locality the potential paradoxes
of which are precisely
avoided by fulfilling~\cite{db2} the MPCP.

\vspace{-.3cm}

\section{General Features and Assumptions of the Model(s)}

\vspace{-.2cm}

The more fundamental gauge group $SMG^3$, because it has one SMG factor
for each of the three generations  of fermions, also
implies three generations of gauge bosons.
As $SMG^3$ is a non-simple group,
it defies
GUT models that espouse
simple groups as the most likely
gauge groups beyond the SMG. Accordingly
$SMG^3$ is referred to as an {\bf A}nti {\bf G}rand
{\bf U}nified {\bf T}heory
(AGUT) gauge group.

Inasmuch as the excellence of our predictions is contingent upon not
encountering new physics in doing the
renormalization group extrapolation
from the Planck scale to that of $M_Z$,
our model does not allow supersymmetry (SUSY).

We predict that the SMG gauge couplings correspond to the point - the
so-called {\em multiple point} - in the phase
diagram of a lattice gauge theory (LGT) with the gauge group $SMG^3$ where the
maximal number of LGT phases convene.
The Planck scale values of the couplings are predicted to be
those of the diagonal subgroup of $SMG^3$ corresponding to the
multiple point (MP)
of $SMG^3$.

The Lagrangian for the gauge group $SMG^3$ has for each of the
SMG subgroups U(1), SU(2) and SU(3)
terms of the form
\begin{equation}
c_{ij}F_{\mu\nu}^iF^{\mu\nu\;j}
\label{gg} \end{equation}
where $i,j \in\{e,\mu,\tau\}$ labels the
generation to which each of the three SMG factors of $SMG^3$ are
associated. Going to the diagonal subgroup, the terms of Eq. \ref{gg}
become identical for each SMG subgroup with the result that the diagonal
subgroup
inverse squared couplings
for the non-Abelian SMG subgroups are enhanced by a factor
$N_{gen}=3$ while that for $U(1)$ is enhanced by roughly a factor
$\frac{1}{2}N_{gen}(N_{gen}+1)=6$ for three generations. This factor 6
arises in the
Abelian case because terms like Eq. \ref{gg} with $i\neq j \in
\{e,\nu,\tau\}$ are allowed for U(1) but forbidden by gauge
invariance for the non-Abelian
SMG subgroups.

The phases we seek are
usually regarded as artifacts of a lattice introduced as a
calculational device.
However, we take a regulator
as an ontological attribute of Planck scale physics inasmuch as
the consistency of all field theories requires
some sort of regulator.
We use a lattice but in
principle it could take other forms (e.g., strings).
In essence, the MPCP leads to finetuning by having multiple
phase coexistence
enforced by having
universally
fixed amounts of extensive quantities. If the interconversion of
such phases occurs by
strongly first order phase transitions, there will be
many combinations of such fixed quantities that can only be
realized by having the coexistence of more than one phase, with the result
that intensive quantities conjugate to the fixed extensive quantities are
fine-tuned.

\vspace{-.3cm}

\section{The Analogy of a Fixed Freezing Point in a Microcanonical Ensemble}

\vspace{-.2cm}

The MPCP can be illustrated by
considering the liquid, solid and gaseous phases of water.
As
transitions between these phases are first order (corresponding to
finite values of the heats of fusion, vapourisation and sublimation),
there is a whole
range of fixed values of the extensive quantities - total energy, volume and
number of molecules of water - that can only be realized by having the
coexistence of all three phases of water. As a result, the intensive
quantities conjugate to energy and volume (i.e., temperature and pressure)
are finetuned to the values at the triple point of water. The triple point
in the phase diagram, spanned by the variables temperature and pressure,
corresponds to the MP in the phase diagram of a LGT.

\vspace{-.3cm}

\section{The Non-abelian Fine Structure Constants from Multiple Point
Criticality}

\vspace{-.2cm}

As the gauge group $SMG^3$ is non-simple,
the idea of a LGT phase as well as the functional
form of the lattice action needs to be generalized
in order to get the LGT phases to
convene at the MP.
It is in the parameter space of such a generalized action that
we seek the MP; the various phases correspond to different regions of this
space.

Assuming a Planck scale lattice as a physical attribute of
fundamental physics implies that
phases distinguishable in the corresponding LGT are
physical and potentially can have physical consequences. Using the
non-simple $SMG^3$ as the gauge group $G$, there are many
distinguishable phases on the lattice, because of the many subgroups and
invariant subgroups of $G=SMG^3$.
In fact, we distinguish one phase for each subgroup $K$ and invariant
subgroup $H$ such that $H \triangleleft  K\subseteq G$.

We classify the various degrees of freedom (DOF) as belonging to the
different phases
according to whether there is spontaneous
breakdown of gauge symmetry (after fixing the gauge appropriately) in the
vacuum under combinations of two types of
gauge transformations. These are global gauge transformations
and local
gauge transformations of the special type that are generated by Lie algebra
vectors that depend linearly on space-time. Denote
these two types of gauge transformations by respectively
$\Lambda_{const}$ and $\Lambda_{lin}$.

Physically, one of our generalized phases corresponds in general
to different patterns of
fluctuation for DOF corresponding to different Lie
sub-algebras. For the phase $(H,K)$ (i.e. $H\triangleleft K\subseteq G$),
the DOF associated with the Lie sub-algebra that
generates the homogeneous space $G/K$ are Higgsed while the subgroup
$K$ corresponds to unHiggsed DOF. The latter DOF have
VEVs that are invariant
under gauge transformations of the type $\Lambda_{const}$.
The DOF with vacuum expectation values (VEVs) that do not
spontaneously break gauge symmetry at all are confined and correspond
to the elements of the maximal invariant subgroup $H$ invariant under
gauge transformations $\Lambda_{const}$ {\em and} $\Lambda_{lin}$.
Coulomb-like DOF are  associated
with the factor group $K/H$ and have VEVs that are invariant under global
gauge transformations $\Lambda_{const}$ but which spontaneously break gauge
symmetry under the
gauge transformations $\Lambda_{lin}$.

We have developed generalised lattice actions that can
provoke confining
phases for almost all the invariant subgroups $H$ of the non-Abelian
subgroups of the SMG
including the discrete subgroups.
Knowing
how to construct approximate phase diagrams for the non-Abelian sector of
the SMG, we can use Monte Carlo results from the literature to
calculate the diagonal subgroup couplings corresponding to the MP
of $SMG^3$. In the case of the non-Abelian couplings, the
continuum inverse squared couplings for $SMG^3$ can essentially be
determined from
the MP of just one
of the SMG factors multiplied by the enhancement factor $N_{gen}=3$
(see comments following Eq. \ref{gg}).
This procedure would be exactly correct if all the invariant subgroups of
$SMG^3$ were of the form of a Cartesian product of factors each of which
stems unambiguously from one of the $SMG^i$ ($i\in \{e,\mu,\tau\}$) in
$SMG^3$.
This is true of Cartesian product subgroups of
$SMG^3$ the factors of which have centers that are disjoint except for the
group identity element. This property, referred to as
factorizability, characterizes almost all of the non-Abelian
subgroups of $SMG^3$.

\vspace{-.3cm}

\section{The More Elusive $U(1)$ Fine Structure Constant Calculation}

\vspace{-.2cm}

Constructing an approximate phase diagram for the Abelian sector of
$SMG^3$ is more complicated than for the non-Abelian DOF,
because all the (infinitely many) subgroups of $SMG^3$ are
{\em invariant} subgroups (since
$U(1)^3$ is Abelian). Each of these can potentially be
made to confine separately from
the rest of the group and therefore should ideally convene at the MP.
As many of these invariant subgroups are
not factorizable in the sense discussed above,
we must deal with the whole Abelian sector
$U(1)^3$  in determining the diagonal subgroup coupling corresponding to the
MP of $SMG^3$.
This latter is our prediction at the Planck scale for the U(1) hypercharge.

In the approximation in which we deal with
just a
single U(1) in determining the Planck scale inverse squared
coupling, the
enhancement factor in going to the diagonal subgroup is more than twice
the factor 3
in the non-Abelian case. This is due to the interactions
between the $U(1)^i$ (with $i\in\{e,\mu,\tau\}$) in
$U(1)^3$. These interactions are related to the fact that,
even in the continuum, terms for which
$i\neq j\in \{e,\mu,\tau\}$  are allowed in Eq. \ref{gg}.

In treating the Cartesian product of $U(1)$ groups, we use a
geometrical description of the rather complicated system of coupling constants
possible for several Abelian gauge fields which can interact via
Lagrangian density terms such as Eq. \ref{gg} with
$i\neq j\in \{e,\mu,\tau\}$.
Basically, this method
uses the square root of the (Manton) plaquette action as the distance
in group space.
Because of the interactions between the U(1) gauge group factors of
$U(1)^3$,
the gauge couplings can be adjusted so as to have hexagonal
symmetry for
the geometry of the identification lattice of points identified
in the covering group $R^3$ upon compactifying to $U(1)^3$.
In this formulation, invariant subgroups correspond to
lines, planes and 3-dimensional figures spanned by the hexagonal
identification lattice points. Hence, if a phase for which an
invariant subgroup $H$ is rendered
confining convenes at the MP, then so do all subgroups into which
$H$ can be rotated under the symmetry operations of the hexagonal
identification lattice. Implementing hexagonal symmetry in
this manner, we get what we
believe to be the maximum number of phases to convene at the MP.

For the important discrete subgroups (i.e. {\bf Z}$_2$ \& {\bf Z}$_3$),
there is some ambiguity as to the extent that the hexagonal symmetry
can be realized. This ambiguity is reflected in the uncertainty of our
prediction for the U(1) coupling.

\vspace{-.3cm}

\section{The Higgs and Top Masses}

\vspace{-.2cm}

The application of the MPCP
to the pure SM with a desert up to the Planck
scale gives a prediction \cite{smtop} of the Higgs boson and top quark
masses, which is
independent of the nature of the fundamental
gauge group at the Planck scale. We use
that the renormalisation group improved
SM effective
Higgs potential $V_{\rm{eff}}(\phi)$ should have two
degenerate minima: $V_{eff}(\phi_{\rm{min}\; 1} = 246 \rm{GeV})  =
V_{eff}(\phi_{\rm{min} \; 2}).$
This means that the vacuum in which we live lies
on the vacuum stability curve and, hence, the Higgs mass
is determined as a function of the top quark mass.
If we use the combined CDF and D0 value for the
top quark pole mass given at this conference,
$M_t = 175 \pm 6$ GeV, the Higgs pole mass is
predicted to be $M_H = 139 \pm 15$ GeV.
However, as pointed out in sections 2 and 3, for the MPCP
to function properly, we require a strongly first
order phase transition. Since we are taking the
Planck scale to be the fundamental scale, this
means that we should take
$\phi_{\rm{min} \; 2} \simeq M_{\rm{Planck}}
\sim 10^{19}\ \rm{GeV}.$
So we obtain a prediction for both the top quark and
Higgs boson pole masses:
{\bf\boldmath $M_{t} = 173 \pm 5\ \mbox{GeV}
\quad M_{H} = 135 \pm 9\ \mbox{GeV}.$}

\vspace{-.3cm}

\section{The Fermion Mass Hierarchy}

\vspace{-.2cm}

In order to understand the other quark and lepton masses,
it is necessary to introduce chiral quantum numbers
beyond those of the SMG. The broken chiral gauge quantum
numbers in our $SMG^3$ model distinguish
between the three generations and have the potential
to generate the fermion mass and mixing hierarchy.
This hierarchy corresponds to different degrees of
suppression for various transitions from right-handed to
left-handed fermion states, each of which carry
different $SMG^3$ gauge quantum numbers. The mass gaps between
the three generations can be explained by suppression
factors due to the
partial conservation of these chiral quantum numbers. However
in order to explain the mass splittings
within each generation (such as the top to bottom
quark mass ratio), it is necessary to extend the $SMG^3$
gauge group by an abelian flavour group factor $U(1)_f$.
The new $U(1)_f$ charges for the fermions are essentially
uniquely determined by the anomaly cancellation conditions.
This $SMG^3\otimes U(1)_f$ model gives the order of magnitude
fit \cite{smg3m}
in Table 2 of the contribution to
these proceedings by Froggatt, Nielsen \& Smith.

\vspace{-.3cm}

\section{The maximal AGUT Group}

\vspace{-.2cm}

It is interesting to note that the $SMG^3\otimes U(1)_f$
Planck scale gauge group, which we have introduced on
phenomenological grounds, can in fact be characterised
as the largest possible AGUT gauge group G satisfying
the following assumptions:
\newline
1. The gauge group G should be a subgroup of the group U(45)
of unitary transformations of the known quark and
lepton Weyl fields (3 generations with 15 in each).
\newline
2. G should {\em not} unite the SMG irreducible
representations, i.e. it should leave the $3 \times 5 =15$
irreducible representations of the SMG as irreducible.
\newline
3. G should have no gauge nor mixed anomalies (and no
Witten anomaly).
\newline
4. G should be the maximal group satisfying the above
conditions.

These assumptions lead to the identification
of G as $SMG^3\otimes U(1)_f$, with the breaking to the
SMG as a diagonal subgroup of G, as we
supposed in the previous sections.

\vspace{-.3cm}

\section{Conclusions}

\vspace{-.2cm}

When the multiple point criticality principle of degenerate vacua
is combined with our maximal AGUT gauge group,
$SMG^3\otimes U(1)_f$, we obtain very good predictions for many
of the parameters of the Standard Model.
It is important for the predictions that the model has
a desert with the Standard Model valid all the way up
close to the Planck scale. In particular the existence of
supersymmetric partners at accessible energies would spoil
our successful predictions of the fine structure constants and the
top quark mass. Apart from the Higgs particle, predicted to
have a mass of 135 $\pm$ 9 GeV, we expect very little new physics
to be found in future generations of accelerators. There are
just a few possibilities:
\newline
1. Flux strings for discrete gauge groups close to the
confinement-Coulomb phase transition.
\newline
2. Scalar particles in the same multiplet as the Weinberg-Salam
Higgs particle.
\newline
3. Particles decoupled from the ones we know.

The principle of degenerate phases is also promising for the solution
of some standard fine-tuning problems. Both a vanishing cosmological
constant, $\Lambda_{cosmological} = 0$, and strong CP conservation,
$\Theta_{QCD} = 0$, are characterised as meeting points of phases
\cite{ambjorn,schierholz}. Unfortunately, however, our hope
of explaining the gauge hierarchy problem, in terms of a
vanishing Higgs mass at a phase boundary,
fails for the strongly first order
phase transition we have argued for in section 6.

Perhaps the most far reaching new physics suggested by
our model is a lack of locality at the
fundamental level, much in the same way as in baby universe
theory. These non-local effects are the same all over space
and time and are precisely of the type which we need to
explain the existence of degenerate vacua. Indeed we seem to
predict the appearance of a new vacuum in the future. This
raises the interesting question of how the critical size
bubble of new vacuum is to be produced and whether human activity is
needed to trigger it off.


\vspace{-.3cm}

\section*{References}

\vspace{-.2cm}

\begin{thebibliography}{99}
\bibitem{db1}
D.L. Bennett and H.B. Nielsen, {\em Intl. J. Mod. Phys. A} (to be
published) hep-ph/9607278

\bibitem{db2} D.L. Bennett (thesis) hep-ph/9607341

\bibitem{db3} D.L. Bennett and H.B. Nielsen, {\em Intl. J. Mod. Phys. A}
{\bf 9} 5155-5200 (1994)

\bibitem{frogniel}
C.D. Froggatt and H.B. Nielsen, {\em Origins of Symmetry} World Science 1991

\bibitem{smtop}
C.D. Froggatt and H.B. Nielsen, \Journal{\PLB}{368}{96}{1996}.

\bibitem{smg3m}
C.D. Froggatt, H.B. Nielsen and D.J. Smith,
\Journal{\PLB}{}{}{\rm{to be published}}, hep-ph 9607250.

\bibitem{ambjorn}
J. Ambj{\o}rn and S. Varsted, \Journal{\NPB}{373}{557}{1992};
\newline
H.W. Hamber, \Journal{\NPB}{400}{347}{1993}

\bibitem{schierholz}
G. Schierholz, {\em Lattice '94, Nucl. Phys.} B,
{\em Proc. Supp.} 42, 270 (1995).

\end{thebibliography}
\end{document}